\documentclass[runningheads]{llncs}
\usepackage[T1]{fontenc}
\usepackage{amsmath}
\usepackage{booktabs}
\usepackage{graphicx}
\begin{document}
\title{Evaluating Theory of Mind and Internal Beliefs in LLM-Based Multi-Agent Systems}
\titlerunning{Evaluating ToM and IB in LLM-Based MAS}
%
\author{Adam Kostka\orcidID{0009-0004-8174-2109} \and\\
Jarosław A. Chudziak\orcidID{0000-0003-4534-8652}}
%
%
\institute{Warsaw University of Technology, Poland
\\
\email{\{adam.kostka.stud,jaroslaw.chudziak\}@pw.edu.pl}}

\maketitle              
\begin{abstract}
LLM-based MAS are gaining popularity due to their potential for collaborative problem-solving enhanced by advances in natural language comprehension, reasoning, and planning. Research in Theory of Mind (ToM) and Belief-Desire-Intention (BDI) models has the potential to further improve the agent's interaction and decision-making in such systems. However, collaborative intelligence in dynamic worlds remains difficult to accomplish since LLM performance in multi-agent worlds is extremely variable. Simply adding cognitive mechanisms like ToM and internal beliefs does not automatically result in improved coordination. The interplay between these mechanisms, particularly in relation to formal logic verification, remains largely underexplored in different LLMs. This work investigates: How do internal belief mechanisms, including symbolic solvers and Theory of Mind, influence collaborative decision-making in LLM-based multi-agent systems, and how does the interplay of those components influence system accuracy? We introduce a novel multi-agent architecture integrating ToM, BDI-style internal beliefs, and symbolic solvers for logical verification. We evaluate this architecture in a resource allocation problem with various LLMs and find an intricate interaction between LLM capabilities, cognitive mechanisms, and performance. This work contributes to the area of AI by proposing a novel multi-agent system with ToM, internal beliefs, and symbolic solvers for augmenting collaborative intelligence in multi-agent systems and evaluating its performance under different LLM settings.

\keywords{Artificial intelligence \and Large language models \and Multi-agent systems \and Theory of Mind \and Internal belief mechanisms \and Logic verification}
\end{abstract}


\begin{figure}[t]
  \centering
  \includegraphics[width=0.8\linewidth]{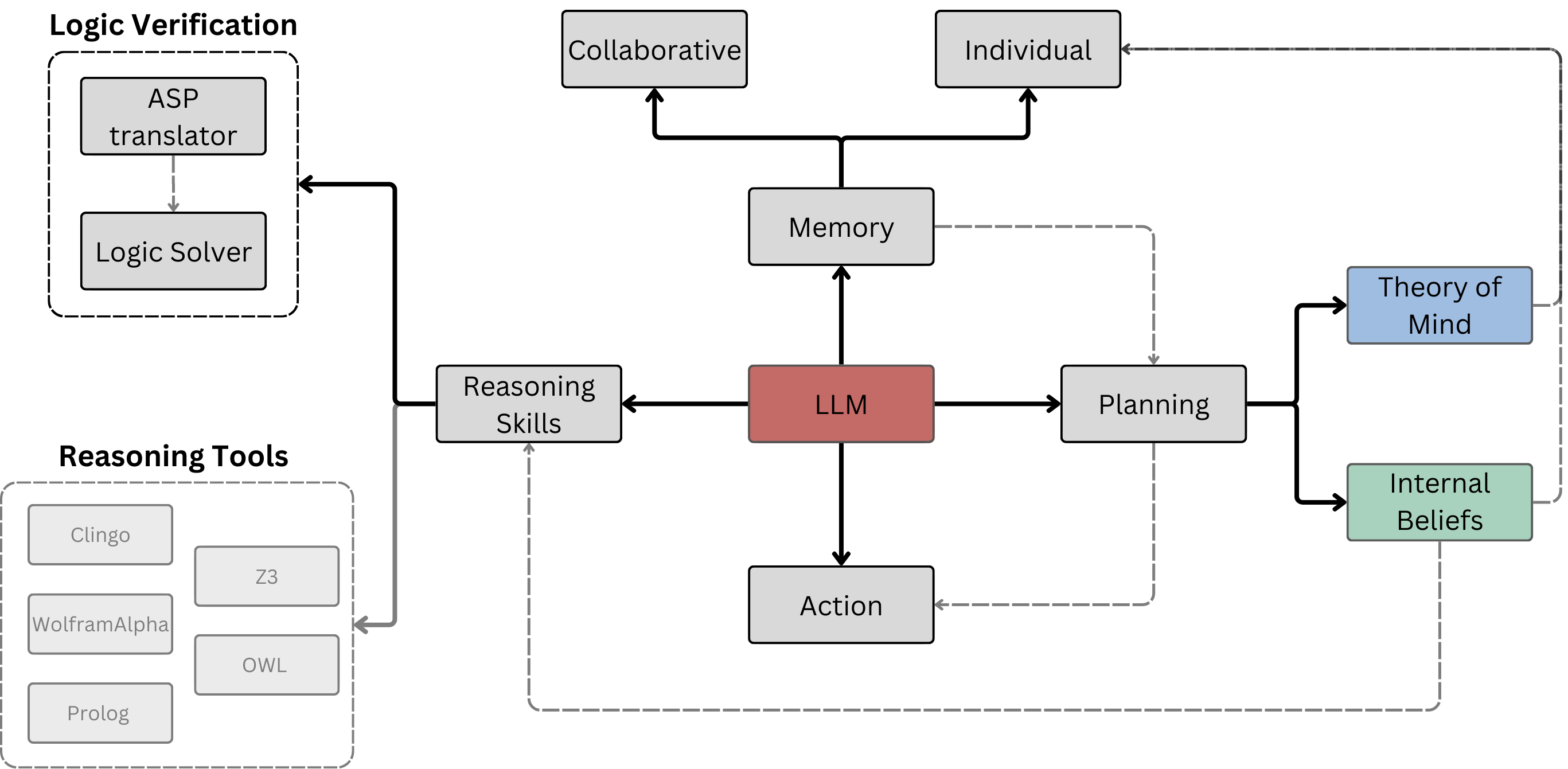}
  \caption{Internal and External Reasoning Processes of LLM Agent (based on \cite{weng2023agent}).}
  \label{fig:agent}
\end{figure}

\section{Introduction}

Multi-Agent Systems (MAS) have been a cornerstone in the resolution of intricate, distributed issues, taking advantage of the strength of coordination between independent agents \cite{Russell_Norvig_2016}. Large Language Models (LLMs) have ushered in a revolutionary age for MAS, providing unmatched possibilities for autonomous, cooperative, and scalable problem-solving in changing situations \cite{ijcai2024p0890,tran2025multiagentcollaborationmechanismssurvey}. LLMs, with their sophisticated natural language understanding and strategic reasoning abilities \cite{naveed2024comprehensiveoverviewlargelanguage}, hold immense promise to revolutionize agent interaction in MAS systems.  In collective intelligence, agents' abilities to infer and act upon the intentions and beliefs of other agents are invaluable. Significantly, research such as \cite{Li_2023} has demonstrated that the integration of Theory of Mind (ToM) serves to enhance collaborative intelligence via agents' capacity for improved prediction and adaptation to their counterparts' objectives. Moreover, work represented by \cite{10.1145/3623809.3623930} has successfully applied the Belief-Desire-Intention (BDI) model to formalize internal belief processes within agent-based systems. Furthermore, the combination of large language models (LLMs) with symbolic solvers, explored in \cite{pan-etal-2023-logic}, is promising for systematic and verifiable reasoning approaches. An example of a system that integrates the above elements is presented in Figure~\ref{fig:agent}.

Even with these developments, few MAS frameworks today utilize these methods. Most modern MAS are based on strict rules and formal communication protocols, which render them incapable of adapting in dynamic and uncertain settings \cite{han2024llmmultiagentsystemschallenges}. While LLMs like ChatGPT, LLama, and Claude have shown strong performance in standalone reasoning and planning tasks, their use in multi-agent settings presents new challenges and opportunities for further development. Effective agent coordination is challenging, particularly with partial or conflicting information. While LLM inference is powerful, it can be inconsistent, leading to suboptimal decisions \cite{chen2024optimaoptimizingeffectivenessefficiency}, especially in situations where speed and accuracy of decision-making are required \cite{synergymas,zhu2025multiagentbenchevaluatingcollaborationcompetition}. Crucially, the synergistic integration of ToM, BDI-based internal beliefs, and symbolic solvers remains underexplored. 

This research directly faces these challenges by raising the following basic questions: How do internal cognitive processes involving symbolic solvers and the Theory of Mind influence collective decision-making in LLM-based multi-agent systems? And how does the interaction among those processes influence the system's accuracy?

For clarity, we outline main concepts pertinent to the present study. Theory of Mind is defined as the capacity of an agent to anticipate the probable actions and intentions of other agents from their assigned roles and observable behavior in the simulation world. Internal Beliefs (IB) are an agent's internal awareness of the game state, represented in a format suitable for logical inference. Logical Verification employs Answer Set Programming (ASP) to check the consistency of an agent's internal beliefs prior to action execution.

This paper makes two main contributions:
\begin{itemize}
    \item We propose a novel agent architecture that integrates ToM, internal belief states, and logical verification using Answer Set Programming to enhance collective intelligence in a multi-agent system.
    \item We carry out a detailed analysis of the impact of different combinations of ToM and internal belief mechanisms on the overall system performance and accuracy.
\end{itemize}
\section{Related Work}
\label{sec:related-work}

Large Language Models (LLMs) in Multi-Agent Systems (MAS) have attracted significant interest and, as a consequence, solutions for cooperative problem-solving \cite{Li_Wang_Zeng_Wu_Yang_2024}.  LLM-based MAS are increasingly applied in decentralized decision-making applications, such as task allocation, resource allocation, and dynamic planning \cite{10.5555/1483085}, as well as in simulating complex human interactions like formal debates \cite{10.1007/978-981-96-6008-7_21}. Several collaborative MAS systems have been developed to solve scalability and coordination problems in distributed systems. For instance, the MegaAgent framework \cite{wang2024megaagentpracticalframeworkautonomous} has demonstrated dynamic task dispatching and efficient use of resources in scalable MAS, while MetaGPT \cite{Hong2023MetaGPTMP} introduced a software programming team based on Standardized Operating Procedures(SOP). In this section, literature pertinent to the topic is summarized around prominent themes: Theory of Mind in MAS, Internal Belief Mechanisms and BDI Models, and LLM-Based Reasoning and Logical Verification.

Theory of Mind (ToM), the cognitive ability to attribute thoughts, desires, and intentions to oneself and others, is crucial for effective social interaction and cooperation \cite{McLaughlin_Beckermann_Walter_2011}. There is growing evidence of emergent ToM capabilities in advanced LLMs like GPT-4 \cite{doi:10.1073/pnas.2405460121}, with metrics like OpenToM being created to measure it \cite{xu-etal-2024-opentom}. Furthermore, ToM enhances collaborative intelligence by enabling agents to better anticipate and respond to others' intentions \cite{Li_2023,10.1007/978-981-96-5881-7_14}. Though distinct from ToM, internal belief mechanisms, codified by the Belief-Desire-Intention (BDI) model \cite{10.1145/1160633.1160820,theBDI,10.1145/3623809.3623930}, are also necessary. These allow agents to maintain and update internal representations of the world and rational action. ToM addresses other's mental states, whereas BDI addresses the agent's internal reasoning process itself. Current ToM MAS research, however, does not fully address integration with internal belief systems and logical verification.

Though LLMs are extremely skilled at natural language processing, their reasoning is error-prone and can be inconsistent \cite{chen2024optimaoptimizingeffectivenessefficiency}. Merging logical tools, e.g., symbolic solvers using Answer Set Programming (ASP) \cite{Heil_2021,Gebser_Kaminski_Kaufmann_Schaub_2013}, provides a formal setting for verification and systematic reasoning. The merging of LLMs with symbolic solvers, as studied in \cite{pan-etal-2023-logic}, opens fascinating avenues for structured and verifiable reasoning. Distributed Default Logic (DDL) \cite{4052922} extends this to distributed reasoning. Also, \cite{yang-etal-2023-coupling} examines how symbolic reasoning crosses over into natural language. 

Despite vast progress, the most crucial issues are still evolving in LLM-based MAS. Logical consistency, memory management, and coordination accomplishment are some of the current research interests. Models like \cite{chen-etal-2024-comm} and \cite{wang2024sibylsimpleeffectiveagent}  are good directions. Explainability and transparency are very critical as well \cite{Calegari_Ciatto_Mascardi_Omicini_2020}. This work draws inspiration from these works but aims to address the core issue of a lack of an integrated multi-agent system where ToM, internal beliefs based on BDI, and logical verification are all integrated together. Our work particularly investigates how these modules cooperate with and affect the behavior of different LLMs under a collective and changing setting. Synergistic integration of such factors is a new path toward enhancing collaborative intelligence in LLM-driven MAS.

\begin{figure}[t]
  \centering
  \includegraphics[width=0.9\linewidth]{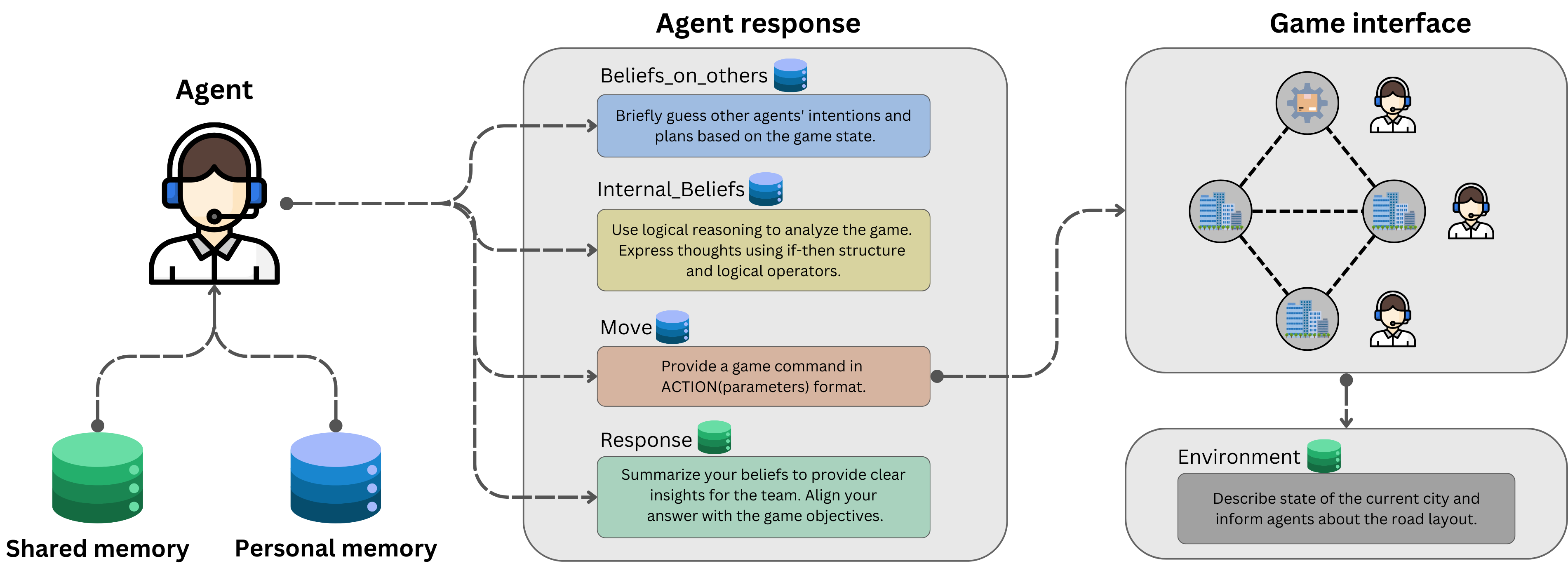}
  \caption{Agents interact in a simulated game using shared and personal memory to form beliefs, make decisions, and coordinate actions based on real-time updates.}
  \label{fig:env}
\end{figure}

\section{Architecture and Task Environment}
\label{sec:architecture}

This section details the architecture of our LLM-based multi-agent system and the virtual environment where the agents interact. The task environment is a dynamic resource allocation problem in which agents must coordinate effectively to maintain the health of urban neighborhoods. We start by describing the task environment, then the agent architecture, and finalize with the integration of logical tools for verification. Figure~\ref{fig:env} displays the agent interaction architecture in the system, showing how individual agents, endowed with shared and private memory, take in information and generate complete responses that interact with a simulated game environment.

\subsection{Task Environment Description}

The task scenario is an interactive city resource allocation simulation designed to challenge the cooperation, decision-making, and resource management skills of a multi-agent system. Each agent is tasked with providing a specific type of resource – FOOD, MEDICINE, or SECURITY – to districts that are facing resource consumption and potential health decline. Since districts require several resources, agents must also share information and coordinate among themselves effectively to meet different needs while managing time, mobility, and capacity limitations. The ultimate aim is to prevent the health of the districts from reaching a critical level by ensuring an appropriate and balanced provision of resources.

The city has four districts, \( D = \{ d_1, d_2, d_3, d_4 \} \), where \( d_1 \) is the supply node, and others are regular districts. Levels of resources, \( R(d, r, t) \), are reevaluated round by round based on consumption by and activity of the agent. The health of districts, \( H(d, t) \), starts at 100 units and decreases in the case of resource shortages. The average health across all non-supply districts at the end of the simulation serves as the primary metric for evaluating system performance.

Agents are capable of performing two actions: MOVE(d) to an adjacent district and SUPPLY\_RESOURCE(x) to transfer resources in the current district. Each agent has a carrying capacity \( C_{\text{max}} \) and starts at \( d_1 \), where resources are resupplied. The consumption rate per district is \( C \) units per turn, and the value is 10. The resource quantities are updated as follows:
\[
R(d, r, t+1) = \max(0, R(d, r, t) - C).
\]
Each turn, a district's health $H(d,t)$ is updated as follows: 
\[
    H(d, t+1) = H(d, t) - \text{Health\_Decrease}(R(d,r,t)),
\]
where the Health\_Decrease is calculated using the following function:
\[
\text{Health\_Decrease}(R(d,r,t)) = 
\begin{cases}
    10 & \text{if } R(d, r, t) < 10, \\
    5  & \text{if } 10 \leq R(d, r, t) < 20, \\
    0  & \text{if } R(d, r, t) \geq 20.
\end{cases}
\]

The districts form a graph \( G(D, E) \) where \( d_1 \) is connected to \( d_2 \) and \( d_3 \), and \( d_2 \) and \( d_3 \) are connected to \( d_4 \). The simulation includes a feedback loop where agents are notified of district health, resources available, and their inventory at the start of each round. If an agent tries to perform an illegal action, like trying to move to a disconnected district, the system provides feedback so that the agent can correct the error.

Agents share information sequentially via a shared memory, which acts as a collective knowledge base for coordination. Private memory (Beliefs about Others, Internal Beliefs, Action) is also stored by each agent in order to enable reasoning at an individual level, separating internal deliberations from team-level communication. Two-level memory allows for scalability and consistency with the potential for independent execution within a collective decision-making system.

\begin{figure}[t]
  \centering
  \includegraphics[width=0.8\linewidth]{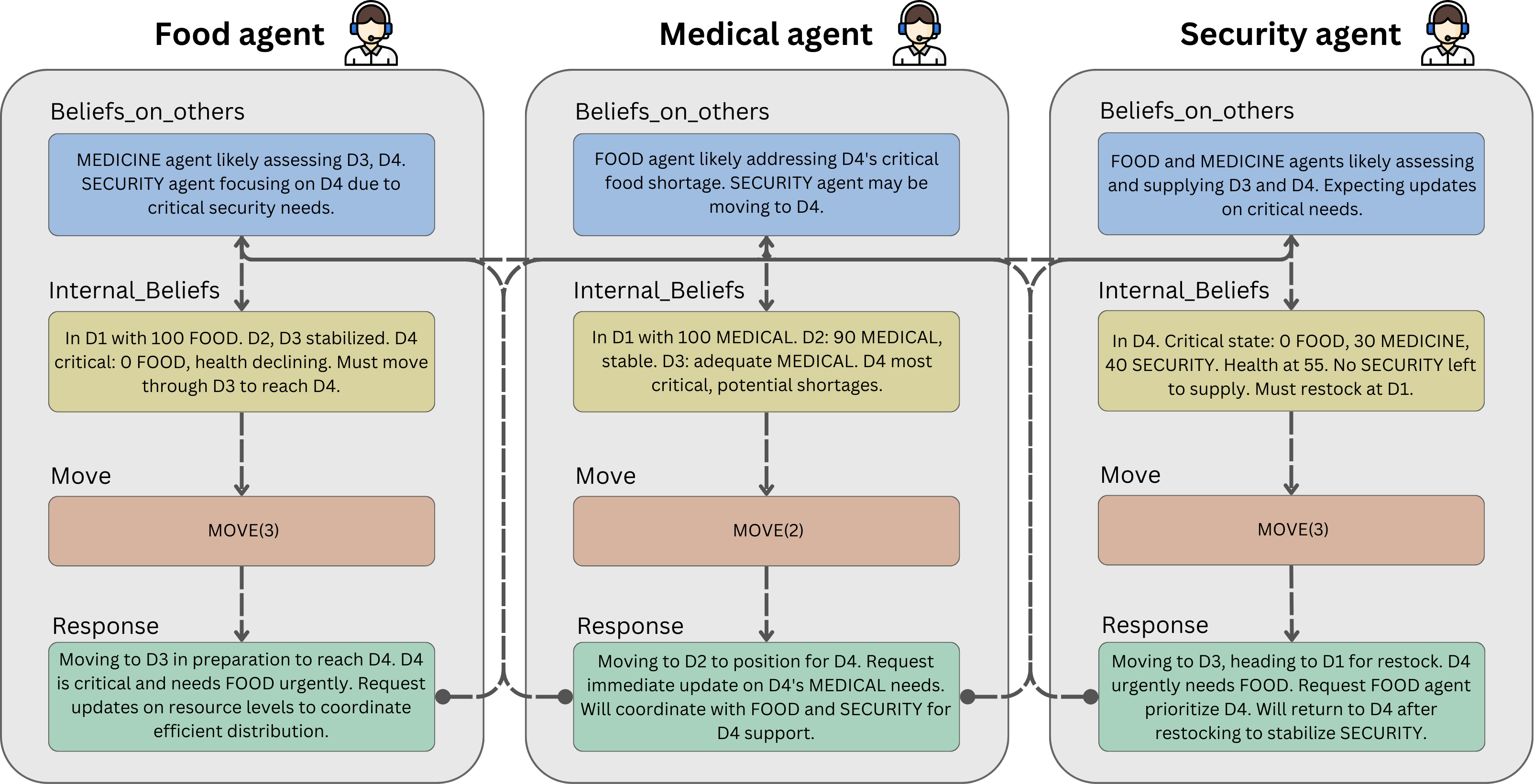}
  \caption{Interaction between Food, Medical, and Security agents showing information flow and exchange of their beliefs, actions, and responses.}
  \label{fig:convo}
\end{figure}

\subsection{Agent Architecture}

Every agent is assigned a specific role within the simulation: FOOD\_AGENT, MEDICAL\_AGENT, or SECURITY\_AGENT. Each agent has an initial prompt stored in its memory. The prompt contains information defining role, task, and environment. It informs the agent of its objectives and the rules of the game simulation. The initial message provides context, for example, the agent's role in dealing with some resources, the importance of cooperation, and the prevention of significant resource shortages. It emphasizes reliance on communication when sharing information on non-observable districts. The prompt instructs the agents to produce answers, neatly specifying the layout, for instance, the Response and Action sections that are always there. If the simulation incorporates ToM or Internal Beliefs, further instructions tell the agent how to reason about peers or internally about the game state. These modular parts can be turned on or off based on the experimental setup. Agents are periodically updated with a summary of the game state, reminding them of goals and reinforcing the necessary response structure.

All agents' responses are crafted into four modular components: Internal beliefs, beliefs about others, Responses, and Actions, as shown in Figure~\ref{fig:convo}. 
The beliefs on Others section reflects the agent's expectation for their fellow agents' likely plans and actions. Utilizing the ToM abilities, the agent reasons over other agents' roles, witnessed activity, and inferred intentions in order to align its strategy with the team action.
Internal Beliefs summarize the internal justification of the agent about the world. Internal Beliefs contain observations, anticipated needs, and plans in logic-friendly representation for computation using the logic tool. Internal beliefs are unshared and used as a basis for the agent's decisions.
The "Beliefs on Others" (ToM) and "Internal Beliefs" (IB) modules interact dynamically. Notably, the ToM module is tasked with developing expectations about the likely actions of other agents. Coupled with the agent's direct observations of the simulated world, i.e., resource levels and the state of the district, these expectations are used in the creation and updating of the internal "Internal Beliefs." The ToM thus tells IB to develop the foundation of the agent's internal reasoning, which is subsequently subjected to consistency checking.
Common to all the agents is the Response field, which contains messages regarding the agent's district at the moment, resource levels, planned actions, and requests for coordination.
The Action attribute specifies the action the agent will execute in the form of, for example, MOVE(district\_number) or SUPPLY\_RESOURCE(amount).

\begin{figure}[t]
  \centering
  \includegraphics[width=1\linewidth]{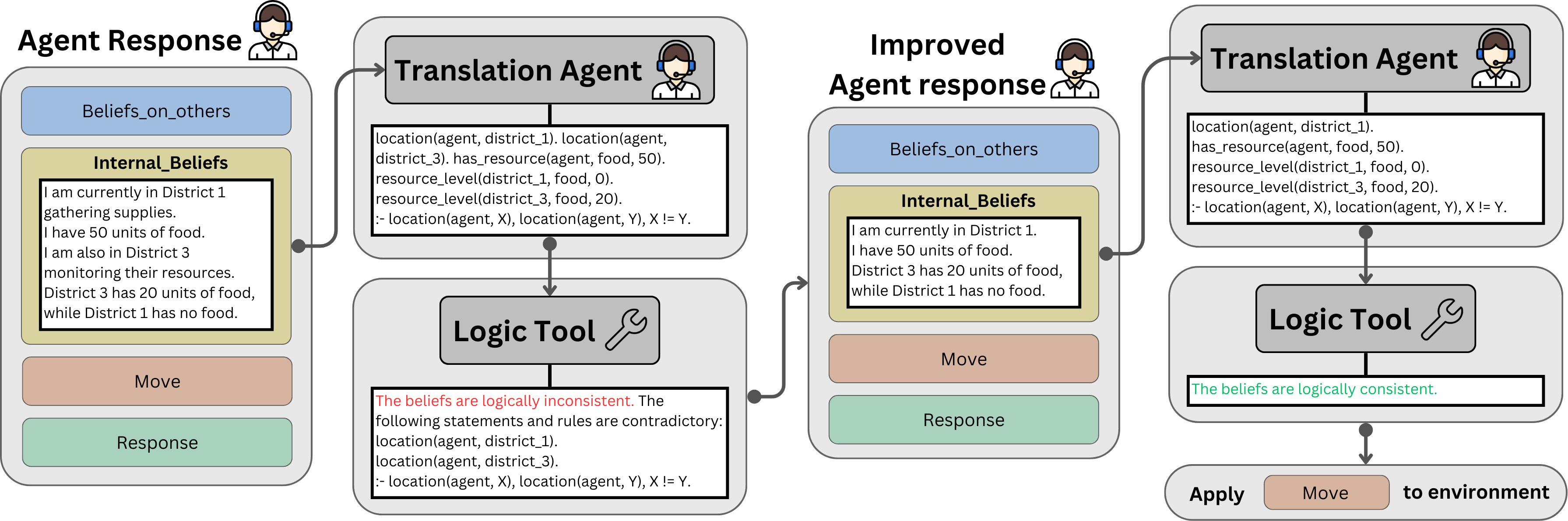}
  \caption{Iterative Logic Verification Pipeline for Cognitive Agents}
  \label{fig:logic}
\end{figure}

\subsection{Logical Tool Integration}
Answer Set Programming (ASP), specifically the Clingo solver, forms the basis for the logical verification of the agent's "Internal Beliefs." ASP is preferred due to its capacity to deal with non-monotonic reasoning, thereby enabling agents to change their beliefs when faced with new facts or remedied contradictions. Its declarative nature also makes it simple and compact to encode both the agent's knowledge and environmental constraints.

The "Internal Beliefs" section is processed through the application of a logic tool derived from Answer Set Programming (ASP). The agent's natural language reasoning is translated into an ASP model that is then checked through the application of the Clingo solver \cite{Gelfond_Kahl_2014}. An illustrative example of this translation process, along with inconsistency detection and correction, is shown in Figure~\ref{fig:logic}. This step ensures the agent's internal reasoning is correct prior to influencing actions. As illustrated in the example, the translation step transforms the agent's natural language statements about its location, resource holdings, and observations of neighboring districts into logical facts and rules of the ASP model. These rules define relations among entities (e.g., agents, districts, resources) and their attributes (e.g., location, amounts of resources). The Clingo solver also checks these rules for consistency with pre-established constraints and with the current knowledge base of the agent and determines potential issues such as logical inconsistency, unsafe variables and cyclic dependencies that may lead to loops.

If inconsistencies are detected, the system generates an explanation to enable the agent to update its beliefs. This explanation process begins by finding the minimal subset of statements within the ASP program, leading to the inconsistency.  The core statements are found using a binary search approach across the assumptions set. Using LLM, the system then translates these elements into natural language, providing feedback about why the contradiction is occurring. This feedback notifies the agent why it holds incoherent beliefs and points it toward forming a logically coherent new set of beliefs. The agent then has the opportunity to reformulate its "Internal Beliefs" based on this feedback and resubmit its response. This iterative process continues for up to three attempts.


\section{Results and Evaluation}
\label{sec:results}

This section presents our experimental findings, comparing the performances of various LLMs in the virtual resource distribution problem. We summarize the experimental setup, present the numerical results in tabular and graphical form, and provide an analysis of agent interaction as well as system performance.

\subsection{Experimental Setup}

Simulations were performed using the multi-agent system outlined in Section 3. The FOOD\_AGENT, MEDICAL\_AGENT, and SECURITY\_AGENT agents worked together to exchange resources across four networked urban districts. The general aim was to prevent districts' health from declining, something that would occur if any one of the three categories of required resources (FOOD, MEDICAL, or SECURITY) were depleted. Each of the three agents was tasked with controlling a separate resource type.

Simulations were run under four distinct configurations:
\\
\textbf{Base Case:} Agents had no access to ToM capabilities ("Beliefs on Others" section) and no internal belief (IB) mechanism ("Internal Beliefs" section).
\\
\textbf{ToM Only:} Agents had access to the "Beliefs on Others" section, allowing them to reason about the likely actions and intentions of other agents.
\\
\textbf{IB Only:} Agents used the "Internal Beliefs" section to maintain an internal representation of their understanding of the game state. This section was also subject to logical verification using Answer Set Programming (ASP).
\\
\textbf{ToM + IB:} Agents had access to both the "Beliefs on Others" and "Internal Beliefs" sections.

The following LLMs were used for testing: ChatGPT-3.5-Turbo, ChatGPT-4o, ChatGPT-4o-mini, Meta Llama 3.1 8B, Claude 3.5 Sonnet.
Every combination of model and configuration was run for multiple trials, each of which was run for up to 7 turns per agent (resulting in 21 agent actions per trial).

It was hypothesized that the integration of ToM and IB mechanisms allowed for agent decision-making and cooperation. ToM, through the "Beliefs on Others" component, should enable reasoning regarding other agents' actions and improve coordination. IB, through the "Internal Beliefs" component and ASP-based verification, should enable deeper and more coherent internal reasoning and offer views on the internal reasoning process of the agent. The principal research question that is addressed is: How do internal belief processes, such as symbolic solvers and Theory of Mind, influence collaborative decision-making in LLM-based multi-agent systems?

\subsection{Results}

\begin{figure}[t]
    \centering
    \includegraphics[width=0.95\textwidth]{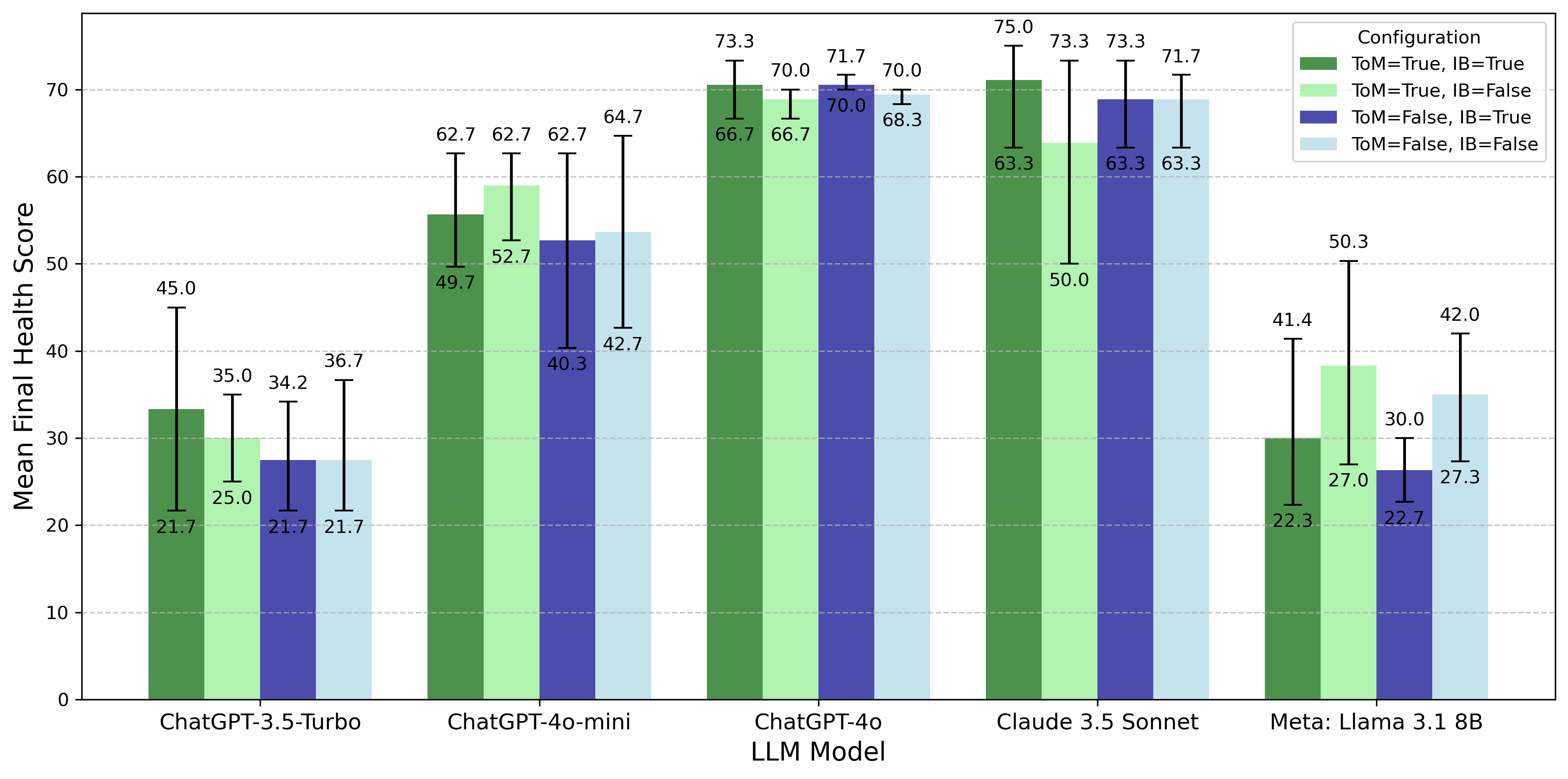}
    \begin{tabular}{lcc|cc}
    \toprule
    LLM Model & \multicolumn{2}{c|}{With ToM} & \multicolumn{2}{c}{Without ToM} \\
    \cline{2-5}
     & With IB & Without IB & With IB & Without IB \\
    \midrule
    ChatGPT-3.5-Turbo & 33.34 & 30.00 & 27.50 & 27.50 \\
    ChatGPT-4o-mini & 55.67 & 59.00 & 52.67 & 53.67 \\
    ChatGPT-4o & 70.56 & 68.89 & 70.56 & 69.44 \\
    Claude 3.5 Sonnet & 71.11 & 63.89 & 68.89 & 68.89 \\
    Meta Llama 3.1 8B & 30.02 & 38.33 & 26.34 & 35.00 \\
    \bottomrule
    \end{tabular}
    \caption{Bootstrapped Median Final Health Scores of districts. The bars represent the median health scores, and the error bars indicate the lower and upper bounds of the 95\% confidence intervals.}
    \label{fig:bootstrapped_results_plot}
\end{figure}

Figure~\ref{fig:bootstrapped_results_plot} presents the median final health scores achieved by each LLM under the different configurations, along with the 95\% confidence intervals (CIs) for these medians, calculated via bootstrapping.

The results reveal significant performance variation between the LLMs and configurations that were tested. ChatGPT-4o consistently reported high health scores in all configurations, with small confidence intervals reflecting high reliability. ChatGPT-3.5-Turbo had lower median health scores in all configurations. ChatGPT-4o-mini performed at intermediate levels, with the scores generally falling between those of ChatGPT-4o and ChatGPT-3.5-Turbo. Meta Llama 3.1 8B's performance varied considerably across configurations, with the best in ToM-only and the worst in no-ToM configurations. Claude 3.5 Sonnet achieved high scores, with its confidence intervals largely overlapping with those of ChatGPT-4o, indicating comparable performance.

\subsection{Analysis of Information Exchange}

The sequential communication protocol had a profound impact on the agents' behavior. Agents not only decided on the present state of the game but also on what they expected other agents to do. Shared memory ("Response" section) was employed to pass plans and observations, while private memory allowed for internal reasoning separation among agents.

The impact of ToM ("Beliefs on Others") was inconsistent. More advanced models sometimes delivered better coordination, successfully predicting the behavior of others. Less capable models often made false predictions, leading to miscoordination, including ToM, which sometimes introduces cognitive load and compromises performance.

The "Internal Beliefs" module, verified by the ASP solver, promoted organized thinking and prevented some logical errors. However, its effectiveness was limited by the vagueness of natural language inputs, though more refined models generated consistent beliefs more successfully.

\subsection{Discussion}

Variability in performance is most likely accounted for by innate LLM properties. ChatGPT-4o being larger was likely responsible for its ability to integrate ToM and IB without loss of performance, which supports greater reasoning ability. ChatGPT-3.5-Turbo's smaller structure may have induced greater cognitive load. The intermediate performance of ChatGPT-4o-mini aligns with the complexity of its architecture. Meta Llama 3.1 8B's variability, particularly with combined ToM+IB, shows insufficient processing for numerous reasoning processes. Claude 3.5 Sonnet performance, as in ChatGPT-4o, indicates a strong architecture for this task.

The results point to the complex interplay between inherent LLM capacities, the inclusion of Theory of Mind (ToM) and Internal Belief (IB) mechanisms, and overall performance in cooperative multi-agent tasks. While ChatGPT-4o excelled in every environment, the other models exhibited significant divergence, suggesting the subtle impact of these cognitive enhancements. 

Interestingly, the inclusion of ToM and IB mechanisms was of a two-edged nature. There were indications of potential gains. For instance, smaller models like ChatGPT-3.5-Turbo showed improvement in performance through the addition of ToM and IB. Similarly, Meta Llama 3.1 8B showed improvement in performance through ToM alone. In some instances, the addition of ToM led to an increase in the upper confidence interval bound of the median health score. All these instances demonstrate that, at best, these mechanisms can enhance coordination and decision-making. The analysis of the interactions among agents found scenarios where accurate prediction of other agents' actions (enabled by ToM) resulted in better resource allocation. The "Internal Beliefs" page, where used effectively, also appeared to promote more structured decision-making.

However, ToM, and IB didn't always boost performance. For some models, engaging these abilities sometimes lowered the CI bound, indicating increased variability. While ToM and IB can assist with coordination, they can cause confusion too, especially for models with more restricted reasoning abilities or in the case of complex scenarios. An LLM might make false assumptions regarding others' intentions or generate inconsistent internal beliefs. The fluctuation in the performance of ChatGPT-4o-mini that was witnessed and the decline in the performance of Meta Llama 3.1 8B upon introducing IB to ToM also validate this, suggesting that the induced cognitive load is contrary at times.

These findings suggest that optimizing the use of ToM and IB in the future would be a good direction for research to pursue, perhaps using more effective prompting methods, hybrid neural-symbolic representations, or learned representations of other agents' behaviors. The inefficiencies inherent in sequential communication protocol evidenced here are also something to remedy; exploring different communication protocols would have a significant effect on information flow. Lastly, the relatively simple nature of this task environment limits the generalizability of these findings.

Overall, this research depicts that while integrating ToM and IB into LLM-based multi-agent systems is extremely promising, the success of such integration relies heavily on the specific LMM, the implementation of the cognitive mechanism, as well as the type of task.


\section{Future Work}
\label{sec:future-work}

Future research should address the enhancement of agent reasoning and coordination within groups, an important problem given the variable impact of the combination of Theory of Mind (ToM) and Internal Belief (IB) mechanisms in various LLMs. These would include investigating more complex ToM applications, like learned models of other agents' goals and behaviors \cite{doi:10.1073/pnas.2405460121,xu-etal-2024-opentom,Li_2023}, more advanced integration of LLMs with symbolic reasoning beyond what is provided by ASP-based verification \cite{pan-etal-2023-logic,yang-etal-2023-coupling,4052922}, and hybrid approaches to reasoning that leverage both neural and symbolic strengths \cite{wang2024openropensourceframework}. Furthermore, future evaluations should incorporate more sophisticated behavioral analysis, such as communication efficiency and planning fallacies, along with formal statistical significance testing, in order to provide a deeper understanding of agent performance.

A prime challenge being most critical is scalability, particularly in terms of task complexity and number of agents. Manifested differences in model performance imply that the incorporation of cognitive mechanisms, in some instances, imposes overhead at the expense of focusing on task-specific reasoning. Future work can tackle the problem of optimizing system architecture and communication protocols to alleviate cognitive overload and optimize coordination efficiency in large-scale systems \cite{qian2024scalinglargelanguagemodelbasedmultiagentcollaboration}. Improved memory management strategies, such as adaptive storage and retrieval, could also be researched for maintaining coherence in long-term interactions \cite{hu2024hiagenthierarchicalworkingmemory}.

The potential applications of this work cover numerous fields. Healthcare, software development, and disaster management models may be enhanced through improved coordination and decision-making  \cite{pandey-etal-2024-advancing,Cinkusz}. Adaptations of this framework may find uses in project management, finance, and other fields that include large amounts of collaboration and higher-order reasoning \cite{DBLP:journals/corr/abs-2407-06567,10.1145/3623762.3633499,paclic_elliottagents}.

Future research should balance computational efficiency with reasoning depth. While Clingo provides logical verification, its computational cost is a constraint; distributed computing or lightweight logic engines offer potential solutions for complex tasks without performance loss \cite{wang2024sibylsimpleeffectiveagent}.

\section{Conclusions}
\label{sec:conclusions}

This paper presented a multi-agent system that sought to enhance cooperation and decision-making in resource allocation scenarios by integrating Large Language Models (LLMs), Theory of Mind (ToM), internal belief mechanisms, and logic-based reasoning modules. The system architecture, such as structured communication protocols and a dual-memory system, facilitated more reflective and coordinated agent interactions. The experiments, which covered a range of LLMs, showed significant variation in performance based on model capacity and the specific integration of cognitive mechanisms.

The key findings of this research demonstrate that the inclusion of ToM and IB mechanisms is not a universally performance-boosting feature for all LLMs. While some models, particularly those with less native reasoning capacity, benefited from these inclusions, others experienced limited or even adverse effects. This shows that LLM choice and careful design of interaction mechanisms are highly significant factors in successful multi-agent cooperation.

This paper presents several important contributions to the field of multi-agent systems. It proposes a new framework for the integration of LLMs, ToM, and logical reasoning and demonstrates the strengths and weaknesses of the approach. It provides a strict empirical evaluation of different LLMs and configurations in a dynamic resource distribution problem and presents some interesting observations about performance-affecting factors. It also identifies key directions for future research, such as building more sophisticated ToM models, exploring hybrid reasoning approaches, improving communication protocols, and enabling more explainability of agent behavior.

The work highlights the need for a multi-aspect approach to building LLM-based multi-agent systems. Simply adding cognitive processes does not guarantee performance growth; particular attention should be given to model selection, task complexity, and interaction mechanism building. With LLMs continuing to evolve, future research must focus on designing robust and resilient frameworks that are able to utilize their strengths efficiently while mitigating their weaknesses within collaborative problem-solving settings. This research provides a foundation for the ongoing exploration of LLMs within multi-agent systems and the evolution of collaborative AI. The presented results provide valuable insights for designing MAS for logistics and resource management scenarios requiring robust collaboration among the agents in the system.


\section*{Acknowledgements}
We would like to acknowledge that the work reported in this paper has been supported in part by the Polish National Science Centre, Poland (Chist-Era IV) under grant 2022/04/Y/ST6/00001

\bibliographystyle{splncs04}
\bibliography{references}
\end{document}